\documentclass[aps,pra,twocolumn]{revtex4}%
\usepackage{amsfonts}
\usepackage{amsmath}
\usepackage{amssymb}
\usepackage{graphicx}
\usepackage{epstopdf}
\usepackage{float}%
\newcommand{\qed}{\nobreak \ifvmode \relax \else
\ifdim\lastskip<1.5em \hskip-\lastskip
\hskip1.5em plus0em minus0.5em \fi \nobreak
\vrule height0.75em width0.5em depth0.25em\fi}

\begin{document}
\title{Gaussian one-way thermal quantum cryptography with finite-size effects}
\author{Panagiotis Papanastasiou}
\affiliation{Computer Science and York Centre for Quantum Technologies, University of York,
York YO10 5GH, United Kingdom}
\author{Carlo Ottaviani}
\affiliation{Computer Science and York Centre for Quantum Technologies, University of York,
York YO10 5GH, United Kingdom}
\author{Stefano Pirandola}
\affiliation{Computer Science and York Centre for Quantum Technologies, University of York,
York YO10 5GH, United Kingdom}

\begin{abstract}
We study the impact of finite-size effects on the security of thermal one-way
quantum cryptography. Our approach considers coherent and/or squeezed states
at the preparation stage, on the top of which the sender adds trusted thermal
noise. We compute the key rate incorporating finite-size effects, and we
obtain the security threshold at different frequencies. As expected
finite-size effects deteriorate the performance of thermal quantum
cryptography. Our analysis is useful to quantify the impact of this
degradation on relevant parameters like tolerable attenuation, transmission
frequencies at which one can achieve security.

\end{abstract}
\maketitle

\section{Introduction}

Quantum key distribution (QKD)~\cite{Gisin2002,Scarani2008} lets two
authorized users (Alice and Bob) to establish unconditionally secure
communication over an insecure quantum channel controlled by an eavesdropper
(Eve). After having shared a secret key, the users can employ it in a one-time
pad protocol. To implement the key distribution, the sender (Alice) sends
non-orthogonal quantum states to the receiver (Bob) through the communication
channel. In this way, the parties can detect Eve's intrusions to gain
information. The absolute privacy of the communication is established
post-processing the raw key by classical protocols of error correction and
privacy amplification, which reduce Eve's information on the final key to a
negligible amount.

Protocols using continuous variable (CV) systems
\cite{Weedbrook2012,diamantiREV} have been proposed for point-to-point one-way
communication, exploiting squeezed states \cite{hillery00,cerf01}, finite
alphabets \cite{norbert02,ralph2007,lev1,christian2017}, Gaussian
\cite{walk013} and non Gaussian post-selection \cite{guo016}. Schemes based on
Gaussian modulations of coherent states have been investigated in great detail
\cite{GG02,grosshans2002reverse,Grosshans2003b,weedbrook2004noswitching,usenko010,usenko1D,ulrik1D}%
, and we have now also experimental implementations over long
distances~\cite{jouguet2013,huang016,zhang017}. Besides one-way protocols, it
has been proposed to exploit two-way communication \cite{pirs2way,
2way2modes,2way2modes2}, quantum illumination \cite{shapiro1}, floodlight QKD
\cite{shapiro2,shapiro3,shapiro4,shapiro5}, and
measurement-device-independence (MDI) \cite{side1,side2}, the latter very
promising to establish end-to-end communications \cite{CV-MDI-QKD,
Ottaviani2015}. In particular CV-MDI protocols are very promising for future
implementation of high-rate metropolitan networks, or for multi-users quantum
conferencing \cite{multikey017}.

Thermal QKD has been investigated in both one-way \cite{weed1, weed2} and
two-way~\cite{weed2014} configuration, with the goal of exploring the
possibility of implementing QKD at frequencies alternative to the optical one.
Initially, the use of thermal states in the optical regime was proposed to
describe imperfections in the preparation of coherent states due to the use of
cheap thermal sources~\cite{filipPRA2008,usenkoREVIEW}. In thermal protocols,
the coherent-state based encoding is replaced by the Gaussian modulation of
thermal states, prepared by adding trusted noise on top of coherent states.
The analysis of the performance at various frequencies is carried out by
expressing the trusted noise in terms of the thermal photon number of the
background radiation.

The increasing attention received by CV-QKD in recent years is justified by
the relative simplicity of the experimental setup, and the very high key-rate
achievable, which can be close to the secret-key capacity of an optical
communication channel, also known as PLOB bound
\cite{pirsPRL2009,PLOB2017,revCH2017}. Moreover, the possibility of
implementing communications exploiting all the electromagnetic spectrum
represents an additional appealing feature of CV systems. The progress
achieved in recent years on the security proofs of Gaussian CV-QKD, has led to
establish composable security proofs for coherent-state one-way
protocols~\cite{Furrer-PRL, leverrierCOMP} and MDI schemes~\cite{cosmoMDI}. An
important scenario to consider, when we study the security of CV-QKD in
practical conditions, is to quantify the security performances of a protocol
when finite-size effects are incorporated in the analysis. The study of
finite-size effects is a precursory step in the security analysis of both
one-way \cite{Leverrierfsz} and MDI schemes \cite{panosMDI}.

Previous studies of thermal protocols have only considered ideal asymptotic
conditions, where the parties exchange infinitely many signals over the
quantum channel. This is a powerful assumption that simplifies the
mathematical complexity of the security analysis: One can work within the
Devetak-Winter security criterion \cite{Devetak207} and use the Holevo
quantity~\cite{holevo1973} to bound Eve's accessible information. The study of
security under more practical conditions requires to assume that Alice and Bob
can only make a finite use of the communication channel. This introduces
finite-size effects that deteriorate the performance, reducing the tolerable
excess of noise, lowering the key rate and shortening the achievable distances.

In this work, we study the impact of finite-size effects on the security of
thermal one-way protocol, adapting the approach described in
Ref.~\cite{ruppertPRA} for coherent state CV QKD. This allows us to quantify
the performance of thermal QKD under more realistic assumptions than in
previous studies. We focus on one-way schemes used in direct reconciliation
(DR) because this represents the configuration providing the best performance
for Gaussian-modulated thermal-state quantum cryptography. The performances
are then limited, by construction, to $3$ dB of channel attenuation.

We systematically analyze the impact of finite-size effects on the performance
of thermal one-way quantum cryptography in various decoding configurations
(homodyne and heterodyne detections), which may be employed in short to
mid-range communication, if one assumes to use optical fibers. Our analysis
also shows that the parameter estimation procedure is negatively affected by
the use of trusted thermal noise, which can further degrade the achievable
distances. We also show that using thermal states, generated starting from
moderately squeezed ones within state-of-the-art experimental equipment (e.g.,
$10$ dB of squeezing), can provide an incremental improvement of the
achievable distance which saturates for higher squeezing factors. Finally, we
study the impact of the finite-size effects on the threshold of a protocol
operating in the microwave regime.

The structure of the paper is the following. Section~\ref{protocol} describes
the protocol, including the optimal attack. In Section~\ref{switching}, we
focus on the case where Bob's decoding is performed by randomly switching the
homodyne detection between the two possible quadratures (switching protocol).
The discussion of other cases (no-switching protocol and encoding based on
squeezed-thermal states, rather than coherent-thermal ones) is given in the
Appendices. In Section~\ref{FS}, we describe the steps to compute the secret
key rate incorporating finite-size effects. In Section~\ref{Results}, we give
the results of our analysis, and discuss the performance of the switching
protocol in terms of the achievable distance in the optical regime, and the
security threshold at various frequencies. Finally, Section~\ref{conclusions}
is left to our conclusions.

\section{\label{protocol}Protocol and eavesdropping}

We now describe the one-way thermal QKD protocol in the prepare and measure
(PM) representation. Additional details on thermal QKD can be found in
Ref.~\cite{filipPRA2008,weed1,weed2,weed2014,ulrik-entropy} and in the recent
review of Ref.~\cite{usenkoREVIEW}. The general bosonic mode of the
electromagnetic field can be described in terms of its quadratures, $Q$ and
$P$, defined as $Q:=a^{\dag}+a$ and $P:=i\left(  a^{\dag}-a\right)  $. We
remark that we assume unit vacuum shot-noise units (SNU) and,
from quadratures $Q$ and $P$ we define the vectorial operator
\[
X:=(Q,P)^{T}.
\]

\begin{figure}[ptb]
\centering\includegraphics[width=0.45\textwidth]{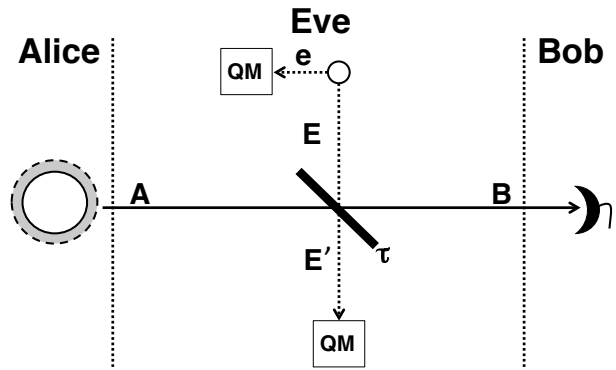}
\caption{The initial mode $A$ is in a thermal state with variance
$V_{\text{th}}+V_{s}$ and modulate dwith variance $V_{M}$. After mode $A$ is
sent through the channel, Bob receives mode $B$ and applies a homodyne
detection on either $q$- or $p$-quadrature or a heterodyne detection measuring
$q$ and $p$-quadrature at the same time. The thermal-loss channel is modelled
by a beam splitter with transmissivity $\tau$. The Gaussian eavesdropping
takes the form of an entangling cloner attack, where Eve's system is described
by the modes $e$ and $E$ in an TMSV state with variance $\omega$. According to
this description, an optical fiber is simulated with transmissivity
$\tau=10^{-\frac{0.2 d}{10}}$, where $d$ (in km) is the length of the fiber,
and excess noise variance $V_{\varepsilon}=\tau\varepsilon$, where
$\varepsilon=\frac{(1-\tau)(\omega-1)}{\tau}$.}%
\label{scheme}%
\end{figure}

The one-way communication goes as follows (see Fig.~\ref{scheme}): Alice
prepares thermal states and modulates them by applying random displacement in
the phase space, according to a bivariate Gaussian distribution. We notice
that the sender can prepare thermal states starting from coherent or squeezed
states. We then have that Alice's input mode, $A$, can be described by the
following input quadrature $X_{A}$%
\begin{equation}
X_{A}=X_{s}+X_{\text{\textrm{th}}}+X_{M}, \label{XA}%
\end{equation}
where $X_{s}$ describes the quantum fluctuations of the initial coherent or
squeezed state from which the sender starts, $X_{\mathrm{th}}$ is the
contribution from trusted thermal noise, while $X_{M}$ describes the Gaussian
encoding. It is easy to see that the resulting input variance, describing
the input mode, is given by the following simple relation%
\begin{equation}
V_{A}=V_{s}+V_{\mathrm{th}}+V_{M},
\end{equation}
where $V_{M}>0$ and $V_{\mathrm{th}}>0$, and $V_{s}=1$ if the sender starts
from coherent states. In the next stage of the protocol, mode $A$ is affected
by a thermal-loss channel. The output mode $B$ is then measured by Bob who can
perform homodyne (switching protocol) or heterodyne detections (no-switching).

The optimal eavesdropping of CV one-way protocols after de Finetti reduction
\cite{cirac-renner, leverrierCOMP} of general attacks, is a single-mode
Gaussian collective attack~\cite{acin07,cerf07,ott17}, completely
characterized in \cite{pirsCAN}. Thermal-loss channels, like free-space and
optical fiber communications, can be dilated into entangling cloners,
consisting of a beam splitter with transmissivity $\tau$, placed between the
parties. This device receives the incoming signal-mode $A$ and Eve's ancillary
mode $E$ (see Fig.~\ref{scheme}). Eve's modes $E$ and $e$ are in a two-mode
squeezed vacuum state (TMSV) $\rho_{eE}$ which is a zero-mean Gaussian state
\cite{Weedbrook2012} described by the covariance matrix (CM)
\begin{equation}
\mathbf{V}_{eE}=%
\begin{pmatrix}
\omega\mathbf{I} & \sqrt{\omega^{2}-1}\mathbf{Z}\\
\sqrt{\omega^{2}-1}\mathbf{Z} & \omega\mathbf{I}%
\end{pmatrix}
, \label{eq:Eve}%
\end{equation}
with variance parameter $\omega\geq1$.
The output modes $E^{\prime}$ and $e$ are then stored in a quantum memory that
is optimally measured by the eavesdropper after the parties have concluded the
communication stage.

In order to quantify Alice-Bob mutual information and Eve's accessible
information, one needs to compute Bob's output mode $B$ (see Fig.~\ref{scheme}%
) which is described by the following vectorial operator
\begin{equation}
X_{B}=(Q_{B},P_{B})^{T}=\sqrt{\tau}(X_{M}+X_{\mathrm{th}}+X_{s})+\sqrt{1-\tau
}X_{0}+X_{\varepsilon}, \label{eq:signal}%
\end{equation}
where $X_{0}$ describes Eve's vacuum mode having variance $V_{0}=1$, and the
term $X_{\varepsilon}$ describes the excess of noise on the channel,
conventionally defined as $\varepsilon:=(1-\tau)(\omega-1)/\tau$ \cite{GG02}.
It is easy to check that the variance of $X_{B}$ can be written as
\begin{equation}
V_{B}=\tau V_{M}+V_{N}, \label{eq:signalvar}%
\end{equation}
with variances $V_{M}\geq0$, $V_{\mathrm{th}}\geq0$, $V_{s}\leq1$, where all
noise contributions are grouped in the term
\begin{equation}
V_{N}=1+V_{\varepsilon}+\tau V, \label{eq:noisevar}%
\end{equation}
and where we have defined the variances of the excess noise as $V_{\varepsilon
}:=\tau\varepsilon$ \cite{ruppertPRA} and $V:=V_{\text{\textrm{th}}}+V_{s}-1$.

\section{\label{switching}Switching protocol with thermal states from
modulated coherent states}

We now consider a specific implementation: Alice starts preparing
Gaussian-modulated coherent states, adds trusted thermal noise, and sends the
resulting signals to Bob who, at random, switches his detection setup between
homodyne measurements on $Q$ or $P$ (switching protocol). We discuss here only
the direct reconciliation (DR), i.e., Bob infers Alice's encoded state from
the outcomes of his detections.

With Alice starting from coherent states, one has the shot-noise variance
$V_{s}=1$, so that $V=V_{\mathrm{th}}$. In such a case, Eq.~(\ref{eq:noisevar}%
) reduces to the simpler expression
\begin{equation}
V_{N}^{c}=1+V_{\varepsilon}+\tau V_{\text{\textrm{th}}}.
\label{eq:noisevar-COH}%
\end{equation}
We notice that, despite DR can only tolerate a maximum of $3~$dB of channel's
attenuation, in case of thermal one-way QKD, it does much better than the RR,
which has been showed to tolerate only a small amount of thermal noise
\cite{weed2014}.

\subsection{\label{mutual info}Mutual information}

From the variances of Eq.~(\ref{eq:signalvar}) and Eq.~(\ref{eq:noisevar-COH}%
), we compute Alice-Bob mutual information%
\begin{equation}
I_{AB}:=H_{B}-H_{B|\alpha}, \label{IAB-GEN}%
\end{equation}
with $H_{B}$ $(H_{B|\alpha})$ being Bob's total (conditional) Shannon entropy
\cite{COVER-THOMAS}. In particular, we may write
\begin{equation}
I_{AB}=\dfrac{1}{2}\log_{2}\dfrac{V_{B}}{V_{B|\alpha}}, \label{eq:mutualinfo}%
\end{equation}
where $V_{B}$ is the variance of Bob's output signal while $V_{B|\alpha}%
=V_{N}^{c}$ is Bob's variance conditioned to Alice's preparation. Therefore,
using Eq.~(\ref{eq:signalvar}) and Eq.~(\ref{eq:noisevar-COH}) we obtain the
following general expression for Alice's Bob mutual information%
\begin{equation}
I_{AB}=\dfrac{1}{2}\log_{2}\left(  1+\dfrac{\tau V_{M}}{1+V_{\varepsilon}+\tau
V_{\mathrm{th}}}\right)  . \label{IAB-switching}%
\end{equation}

\subsection{\label{Secret key rate}Key rate}

Under ideal conditions of infinite number of channel uses, we can write the
Devetak-Winter rate \cite{Devetak207}
\begin{equation}
R:=I_{AB}-I_{E}, \label{ideal-KEY}%
\end{equation}
where Eve's accessible information, $I_{E}$, is computed with the Holevo
function \cite{holevo1973}. In DR the quantity $I_{E}$ is given by
\begin{equation}
I_{E}=S_{eE^{\prime}}-S_{eE^{\prime}|\alpha}, \label{IEDR}%
\end{equation}
where $S_{eE^{\prime}}$ and $S_{eE^{\prime}|\alpha}$ describe the total and
conditional von Neumann entropies of the output states $\rho_{eE^{\prime}}$
and $\rho_{eE^{\prime}|\alpha}$. For Gaussian states, the von Neumann
entropies are completely determined by their CMs $\mathbf{V}_{eE^{\prime}}$
and $\mathbf{V}_{eE^{\prime}|\alpha}$ taking the following simple form
\cite{Weedbrook2012}
\begin{equation}
S=\sum_{i}h(\nu_{i}), \label{von-neumann}%
\end{equation}
where the entropic function $h(.)$ is defined as
\begin{equation}
h(x):=\frac{x+1}{2}\log_{2}\frac{x+1}{2}-\frac{x-1}{2}\log_{2}\frac{x-1}{2},
\label{h-func}%
\end{equation}
and $\nu_{i}$ are the corresponding symplectic
eigenvalues~\cite{Weedbrook2012}.

Moving from ideal conditions to realistic scenarios, the parties extract a
usable key from a finite number of uses of the quantum channel. This generally
deteriorates the performances because the efficiency of the classical
protocols of error correction and privacy amplification is reduced, as well as
the accuracy of the channel parameter estimation. A first adjustment to the
key-rate of Eq.~(\ref{ideal-KEY}) incorporates the efficiency of classical
protocols, and is given by the following key rate%
\begin{equation}
R_{\xi}=\xi I_{AB}-I_{E}, \label{R-beta}%
\end{equation}
with efficiency $\xi\leq1$. We remark that the design of efficient classical
error correction codes, such that $\xi\simeq1$ is non-trivial, but recent
progress~\cite{MILICEVIC,joseph} showed that efficiencies as large as
$\xi\simeq0.98,$ or more, are achievable today. For this reason this
imperfection should not be considered as a major bottleneck for the
development of CV quantum cryptography.

\section{\label{FS}Finite-size description}

The key rate $R_{\xi}$ of Eq.~(\ref{R-beta}) clearly fails to intercept all
finite-size effects which play a role in quantifying the parameters of the
attack which, accordingly to the discussion in Sec.~\ref{protocol}, is
quantified by excess of noise $V_{\varepsilon}$ and transmissivity $\tau$. In
this section, we quantify the impact of finite-size effects by adapting the
approach described in Ref.~\cite{ruppertPRA}, which is fairly simple to
generalize to the thermal case. We can define two statistical variables
$M_{i}$ and $B_{i}$, for $i=1,\dots,m$, representing the realizations of the
input $X_{M}$ and of the output mode $X_{B}$ of Eq.~(\ref{eq:signal}). The
definition of the estimator of covariance $\hat{\sigma}_{MB}$, between modes
$X_{M}$ and $X_{B}$, is then easy to define as follows
\begin{equation}
\hat{\sigma}_{MB}=\frac{1}{m}\sum_{i=1}^{m}M_{i}B_{i}. \label{Cmb}%
\end{equation}
From Eq.~(\ref{Cmb}) we can compute both expectation value and variance.
Assuming $M_{i}$ and $B_{i}$ as independent and normally distributed Gaussian
variables, we get the expectation value%
\begin{equation}
\mathbb{E}\left[  \hat{\sigma}_{MB}\right]  =\sqrt{\tau}V_{M}=\sigma_{MB},
\label{E-sigma}%
\end{equation}
and the variance%
\begin{equation}
V_{\hat{\sigma}_{MB}}=\frac{\tau V_{M}^{2}}{m}\left(  2+\frac{V_{N}}{\tau
V_{M}}\right)  . \label{V-sigmaMB}%
\end{equation}
Similarly, we can obtain expectation value and variance of the estimator,
$\hat{\tau}$, of the transmissivity $\tau$. From Eq.~(\ref{E-sigma}), one then
writes%
\begin{equation}
\hat{\tau}=\frac{\hat{\sigma}_{MB}^{2}}{V_{M}^{2}}=\frac{V_{\hat{\sigma}_{MB}%
}}{V_{M}^{2}}\left(  \frac{\hat{\sigma}_{MB}}{\sqrt{V_{\hat{\sigma}_{MB}}}%
}\right)  ^{2}, \label{tau-estimator}%
\end{equation}
where $\left(  \frac{\hat{\sigma}_{MB}}{\sqrt{V_{\hat{\sigma}_{MB}}}}\right)
^{2}$ is chi-squared distributed.

From Eq.~(\ref{tau-estimator}), we can compute the following expectation
value
\begin{equation}
\mathbb{E}\left[  \hat{\tau}\right]  =\frac{V_{\hat{\sigma}_{MB}}}{V_{M}^{2}
}\mathbb{E}\left[  \left(  \frac{\hat{\sigma}_{MB}}{\sqrt{V_{\hat{\sigma}
_{MB}}}}\right)  ^{2}\right]  =\tau+\mathcal{O}(1/m), \label{AVERAGE-tau}%
\end{equation}
having confidence interval quantified by variance
\begin{equation}
\sigma_{\hat{\tau}}^{2}=\frac{4\tau^{2}}{m}\left(  2+\frac{V_{N}}{\tau V_{M}
}\right)  +\mathcal{O}(1/m^{2}). \label{VAR-tau}%
\end{equation}

The same steps can be made to obtain the variance $V_{N}$ starting from the
statistical sampling $B_{i}$ and $M_{i}$. Using Eq.~(\ref{eq:signalvar}) we
can write the estimator $\hat{V}_{N}$ as follows
\begin{equation}
\hat{V}_{N}=\dfrac{1}{m}\sum_{i=1}^{m}\left(  B_{i}-\sqrt{\hat{\tau}}%
M_{i}\right)  ^{2}. \label{VN-EST}%
\end{equation}
It is clear from Eq.~(\ref{AVERAGE-tau}) and Eq.~(\ref{VAR-tau}) that the
standard deviation $\sigma_{\hat{\tau}}$ becomes rapidly negligible as $m\gg
1$. One can then safely replace the estimator $\hat{\tau}$ with its actual
value $\tau$ in Eq.~(\ref{VN-EST}). Then, noticing that variable $B_{i}%
-\sqrt{\tau}M_{i}$ is normally distributed with variance $V_{N}$, we have that
$\sum_{i=1}^{m}\left(  \frac{B_{i}-\sqrt{\tau}M_{i}}{\sqrt{V_{N}}}\right)
^{2}$is also $\chi^{2}$-distributed with expectation values $m$ and variance
$2m$. We then can write
\[
\hat{V}_{N}=\frac{V_{N}}{m}\sum_{i=1}^{m}\left(  \frac{B_{i}-\sqrt{\tau}M_{i}%
}{\sqrt{V_{N}}}\right)  ^{2}.
\]
The estimator for the variance $V_{\varepsilon}$, can now be expressed using
$\hat{V}_{N}$ and $\hat{\tau}$. It is easy to check that one obtains the
following formula
\[
\hat{V}_{\varepsilon}=\hat{V}_{N}-\hat{\tau}V-1,
\]
with expectation value
\begin{equation}
\mathbb{E}(\hat{V}_{\varepsilon})=V_{N}-\tau V-1, \label{AVERAGE-Veps}%
\end{equation}
and variance
\begin{equation}
\sigma_{\hat{V}_{\varepsilon}}^{2}=\frac{2V_{N}^{2}}{m}+V^{2}\sigma_{\hat
{\tau}}^{2}. \label{S-switch}%
\end{equation}
We remark that these equations are formally identical to the case described in
Ref.~\cite{ruppertPRA}. The only but crucial difference, in our case, is the
presence of the contribution from thermal noise $V_{\mathrm{th}}$, which
appears in $V$.

Assuming an error probability for the parameter estimation of the order of
$\varepsilon_{PE}=10^{-10}$, we can associate confidence intervals of
$6.5$-sigmas which allow us to write the values of transmissivity and excess
noise as%
\begin{equation}
\tau^{\text{\textrm{low}}}:=\hat{\tau}-6.5\sigma_{\hat{\tau}}(\hat{\tau}%
,\hat{V}_{\varepsilon})\text{,~}V_{\varepsilon}^{\text{\textrm{up}}}:=\hat
{V}_{\varepsilon}+6.5\sigma_{\hat{V}_{\varepsilon}}(\hat{\tau},\hat
{V}_{\varepsilon}). \label{eq:confidencesw}%
\end{equation}
The quantities in Eq.~(\ref{eq:confidencesw}) are then used to compute the
finite-size key rate, which is given by the following expression%
\begin{equation}
K=\frac{n}{\bar{N}}\left[  R_{\xi}(\xi,V_{s},V_{M},V_{\text{\textrm{th}}%
},V_{\varepsilon}^{\text{\textrm{up}}},\tau^{\text{\textrm{low}}}%
)-\Delta\right]  , \label{eq:finite size secret key rate}%
\end{equation}
where $\bar{N}=n+m$, is the total number of signals points, $n$ is the number
of signals used to build the key, and the correction term $\Delta$ accounts
for the penalty for using the Holevo bound in the key rate of
Eq.~(\ref{eq:finite size secret key rate}) using a finite number of signals.
(Its description can be found in ~\cite{Leverrierfsz,ruppertPRA}).

\section{\label{Results}Performances and discussion}

In this section, we discuss the performance of finite-size thermal one-way QKD
DR. The results are obtained numerically, evaluating the key rate of
Eq.~(\ref{eq:finite size secret key rate}), and quantifying relevant
quantities like achievable distance, block-size dimensions needed to obtain a
positive key or to recover the asymptotic key rate, and the finite-size
performances of thermal QKD at different frequencies. \begin{figure}[ptb]
\centering\includegraphics[width=0.55\textwidth]{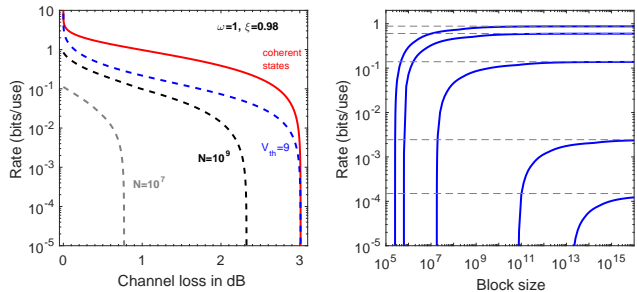}\caption{(Color
online) This figure focuses on the key-rate in the optical regime. The left
panel describes the key rate versus channel attenuation given in dB. The red
solid curve describes the ideal key-rate, using just coherent states. The
blue-dashed curve describes the ideal key rate assuming a preparation noise
with variance $V_{\mathrm{th}}=9$ SNU. Then, we keep the same $V_{\mathrm{th}%
}$ and plot the finite-size rate for block size with $\bar{N}=10^{9}$ (black
dashed line) and $\bar{N}=10^{7}$ (gray dashed) with $\xi=0.98$ and $\omega
=1$. The key rate is optimized over the Gaussian modulation $V_{M}$. The right
panel presents the key rate as a function of the block size ($\bar{N}$). We
fix channel attenuation to $1$ dB, and we assume pure loss attack $\omega=1$
while $\xi=0.98$. The plot shows the convergence of the key rates toward the
asymptotic values (dashed curves) for different values of the preparation
noise $V_{\mathrm{th}}=$ $0$,$~1,~10,~100,~150$ SNU, from top to bottom.}%
\label{SUM-RATE}%
\end{figure}

\subsection{Secret key rate for different block sizes in the optical regime}

Here we focus on the size of the signal blocks needed in order to achieve a
positive key rate in the presence of increasing thermal noise. We use the
average values $\langle\hat{\tau}\rangle\simeq\tau$ and $\langle\hat
{V}_{\varepsilon}\rangle=V_{\varepsilon}$ for which one can write
$V_{\varepsilon}=\tau\varepsilon$, with $\varepsilon=\left[  (1-\tau
)\omega-(1-\tau)\right]  /\tau$. The parameter $\omega$ represents the
variance of thermal noise of Eve's ancillary states used in the attack. We
write the transmissivity $\tau$ in terms of $dB$ of attenuation defining
$\tau=10^{-\frac{dB}{10}}$ and we express the key rate as follows%
\begin{equation}
K=(1-r)\left[  R_{\xi}(\xi,V_{s},V_{M},V_{\text{\textrm{th}}},\omega
,dB,r,\bar{N})-\Delta\right]  , \label{eq:finitesizeeffective}%
\end{equation}
where $r:=m/\bar{N}$. From Eq.~(\ref{eq:finitesizeeffective}) we can plot the
key rate as a function of the channel attenuation, fixing the values of
$V_{\mathrm{th}}$, efficiency $\xi$, thermal noise $\omega$, and shot-noise
variance $V_{s}$. Then we can optimize over the remaining parameters. The
results for pure-loss attacks are shown in Fig.~\ref{SUM-RATE}. In the left
panel we plot the key rate for different values of the block-size and
preparation noise. In particular, the red solid line describes the asymptotic
key rate when Alice send coherent states, i.e., $V_{\mathrm{th}}=0$, while the
blue-dashed line is for $V_{\mathrm{th}}=9$ SNU. \begin{figure}[ptb]
\centering\includegraphics[width=0.4\textwidth]{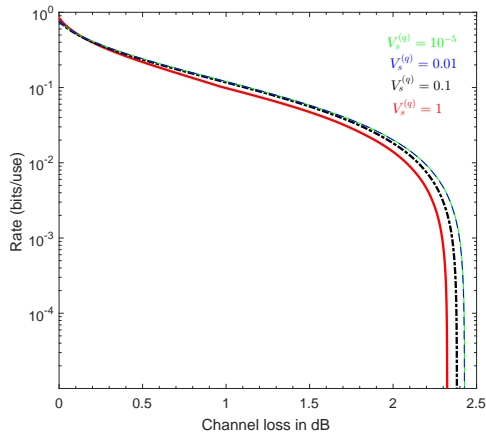}.\caption{(Color
online) We consider the case for signal points block-size of $\bar{N}$
$=10^{9}$, reconciliation efficiency $\beta=98\%$, trusted thermal noise of
$V_{\mathrm{th}}=9$ SNU and pure loss attack $\omega=1$. We compare the
finite-size key rate obtained when Alice starts from coherent states
($V_{s}^{(q)}=V_{s}^{(p)}=1$) with the case when $V_{s}^{(q)}=10^{-1}$, i.e.,
Alice's encoding is performed by adding thermal noise on moderate squeezed
states. We notice that in such a case the achievable distance is only
incrementally improved. Moreover the performance of the protocol saturates for
stronger initial squeezing, i.e. for $V_{s}^{(q)}=10^{-2}$ (blue-dashed),
$10^{-5}$ (green line).}%
\label{plot-squeezing}%
\end{figure}Then, we compare the previous curves with the key rate of
Eq.~(\ref{eq:finitesizeeffective}) for $\bar{N}=10^{9}$ (black dashed line)
and $\bar{N}=10^{7}$ (gray dashed line).

In Fig.~\ref{SUM-RATE} (right panel), we quantify the block-size needed to
achieve a positive key rate for increasing values of the preparation noise. We
fix the attenuation to $1$ dB and assume pure loss attack ($\omega=1$ SNU). We
then plot the key-rate as a function of the block-size, for preparation noise
$V_{\mathrm{th}}=0$,$~1,~10,~100,~150$ SNU from top to bottom and efficiency
$\xi=0.98$~\cite{MILICEVIC}. Our results show that, by an increase in
$V_{\mathrm{th}}$, the block-size need to be increased in order to match the
asymptotic value of the key rate (dashed lines).

Finally Fig.~\ref{plot-squeezing} compares the key rate of the switching
protocol when Alice start from coherent states (red solid line) with the case
where she start from squeezed states. To distinguish between these two cases
Eq.~(\ref{eq:noisevar}) splits as follows%
\begin{align}
V_{N}^{(q)}  &  =1+V_{\varepsilon}+\tau\left(  V_{th}+V_{s}^{(q)}-1\right)
,\label{NSQUEEZING-q}\\
V_{N}^{(p)}  &  =1+V_{\varepsilon}+\tau\left(  V_{th}+V_{s}^{(p)}-1\right)  ,
\label{NSQUEEZING-p}%
\end{align}
where $V_{s}^{(p)}=1/V_{s}^{(q)}$. For coherent states $V_{s}^{(q)}%
=V_{s}^{(p)}=1$ and we recover Eq.~(\ref{eq:noisevar}). This case is described
by the red line in Fig.~\ref{plot-squeezing}, while the others lines describe
the cases $V_{s}^{(q)}=10^{-1}$ SNU (black dot-dashed), $V_{s}^{(q)}=10^{-2}$
(blue-dashed) and $V_{s}^{(q)}=10^{-5}$ (green). We see that using squeezed
states can only incrementally increase the achievable distances, which
saturates as the degree of squeezing increases.


\subsection{\label{Rate at various electromagnetic wavelengths} Security
thresholds with finite-size effects at different frequencies}

In order to study the performance of the protocol at different frequencies, we
follow the approach used in~\cite{weed1,weed2014}. We rewrite the preparation
noise variance $V_{\text{\textrm{th}}}\geq0$, as%
\begin{equation}
V_{\text{\textrm{th}}}=2\bar{n}, \label{Vth-nBAR}%
\end{equation}
where the average thermal photon number $\bar{n}$ is given by the Planck's
formula%
\[
\bar{n}=\frac{1}{\exp\left(  \frac{hf}{k_{B}T}\right)  -1},
\]
at temperature $T$. The quantity $h$ is Planck's constant, $k_{B}$ is
Boltzmann's constant, and $f$ represents the frequency of the signals.

\begin{figure}[ptb]
\centering\includegraphics[width=0.4\textwidth]{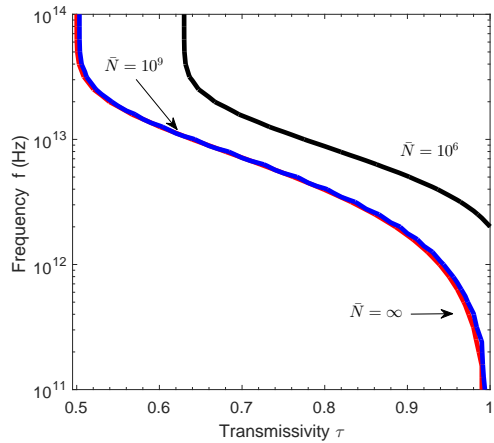}\caption{(Color
online) The red line shows the security threshold (frequencies vs channel's
transmissivity) for a shot-noise level attack $\omega=V_{th}+1$ without
finite-size effects, and assuming infinite Gaussian modulation. Then we have
the case for block size with $\bar{N}=10^{6}$ signal points (black) and
$\bar{N}$ $=10^{9}$ (blue).}%
\label{summary}%
\end{figure}

Therefore, the shot-noise level of Bob's detectors operating in the microwave
regime will be different from the shot noise level in the optical regime,
which is equal to $1$ in vacuum shot-noise units. This shot noise will be
given with respect to Alice's thermal mean photon number and will lead to an
entangling cloner attack with $\omega=V_{\mathrm{th}}+1$. Assuming room
temperature of $T=300$ Kelvin and replacing $\omega=V_{\mathrm{th}}+1$, we can
rewrite the key rate $K$ as function of frequency $f$ and transmissivity
$\tau$. The corresponding threshold of the rate for different block sizes is
illustrated in Fig.~\ref{summary} and shows that, in the microwave region,
security is achieved only for transmissivities very close to $\tau=1$ for a
moderately high block size number of $N=10^{9}$. 

\section{Conclusion\label{conclusions}}

In this work, we studied the security of thermal one-way quantum cryptography,
including finite-size effects. These are evaluated adapting the estimation
theory developed in Ref.~\cite{ruppertPRA} \ suitably extended to the case of
thermal protocols. We focused on the protocol used in direct reconciliation
because it is known that one-way protocols in reverse reconciliation cannot
work at micro-wave frequencies.

Our analysis confirms that implementing CV-QKD with Gaussian modulated thermal
states is challenging, and we cannot achieve long distance communications when
we move away from a pure-loss attack scenario. When thermal noise increases
(for instance $V_{\mathrm{th}}>10$) both key rate and achievable distance
rapidly deteriorate. This is caused by the role of the preparation noise
variance $V_{\mathrm{th}}$ on the confidence interval. In fact, the use of
large amount of trusted noise, spreads the confidence intervals reducing the
transmissivity and increasing the noise to be considered. This determines a
degradation of the performance, which can only be balanced by increasing the
block-size. This degradation rapidly worsening when the protocol is operating
in the microwave regime since in such a case the typical detector's shot-noise
implies an entangling cloner attack with too high thermal noise.

Finally we remark that alternative approach, based on schemes exploiting
post-selection and two-way communication might be more effective in the
thermal regime. This will be investigated in future works.

\section{Acknowledgements}

This work has been supported by the EPSRC via the `UK Quantum Communications
HUB' (Grant no. EP/M013472/1). Authors acknowledge V. Usenko for feedback.

\appendix

\section{No-switching protocol\label{No-switching}}

In this appendix, we focus on the no-switching protocol studying both DR and
RR.
The description of the statistical estimators for the no-switching protocol is
clearly analogous to that described in the main text for the switching
protocol. For the no-switching scheme, we build two estimators, one for each
quadrature $q$ and $p$. The optimal estimators of transmissivity and excess of
noise are then computed by combining them in the optimal linear combination.

Let $B^{\prime}$ describing Bob's output after the fifty-fifty beam splitter.
The vectorial quadrature $X_{B^{\prime}}=(q_{B^{\prime}},p_{B^{\prime}})^{T}$
has entries given by%
\begin{align}
q_{B^{\prime}}  &  =\dfrac{q_{B}+q_{\text{\textrm{vac}}}}{\sqrt{2}},\\
p_{B^{\prime}}  &  =-\dfrac{p_{B}-p_{\text{\textrm{vac}}}}{\sqrt{2}},
\end{align}
where $q_{\text{\textrm{vac}}}$ and $p_{\text{\textrm{vac}}}$ describe the
contributions from the vacuum mode mixed with mode $B$ at the final beam-splitter.

From Eq.~(\ref{eq:signalvar}), one can write the variances of mode $B$ as
follows%
\begin{align}
V_{B}^{q}  &  =\tau V_{M}+V_{N}^{q},\\
V_{B}^{p}  &  =\tau V_{M}+V_{N}^{p}.
\end{align}
In the general case, where Alice starts from squeezed states, the noise
contributions are given by the expressions%
\begin{align}
V_{N}^{q}  &  =1+V_{\varepsilon}+\tau(V_{\mathrm{th}}+V_{s}-1),\\
V_{N}^{p}  &  =1+V_{\varepsilon}+\tau(V_{\text{\textrm{th}}}+1/V_{s}-1).
\end{align}
One can write the following output quadratures of mode $B^{\prime}$%
\begin{align}
q_{B^{\prime}}  &  =\sqrt{\dfrac{\tau}{2}}q_{M}+q_{N^{\prime}},\\
p_{B^{\prime}}  &  =-\left(  \sqrt{\dfrac{\tau}{2}}p_{M}+p_{N^{\prime}%
}\right)  ,
\end{align}
where $q_{N^{\prime}}$ and $p_{N^{\prime}}$ are given by%
\begin{align}
q_{N^{\prime}}  &  =\dfrac{1}{\sqrt{2}}\left(  q_{N}+q_{\text{\textrm{vac}}%
}\right) \label{eq:qunprime}\\
p_{N^{\prime}}  &  =\dfrac{1}{\sqrt{2}}\left(  p_{N}-p_{\text{\textrm{vac}}%
}\right)  .
\end{align}
These have variances%
\begin{align}
V_{B^{\prime}}^{q}  &  =\dfrac{1}{2}\tau V_{M}+V_{N^{\prime}}^{q},\\
V_{B^{\prime}}^{p}  &  =\dfrac{1}{2}\tau V_{M}+V_{N^{\prime}}^{p},
\end{align}
and where%
\begin{align}
V_{N^{\prime}}^{q}  &  =\dfrac{V_{N}^{q}+1}{2}=\dfrac{2+V_{\varepsilon}%
+\tau(V_{\text{\textrm{th}}}+V_{s}-1)}{2}\label{eq:noisevarnswq}\\
V_{N^{\prime}}^{p}  &  =\dfrac{V_{N}^{p}+1}{2}=\dfrac{2+V_{\varepsilon}%
+\tau(V_{\text{\textrm{th}}}+1/V_{s}-1)}{2}.
\end{align}
If Alice uses coherent states $V_{s}=1$, the previous formulas simplify to the
following expressions%
\begin{equation}
V_{N^{\prime}}^{c}=V_{N^{\prime}}^{q}=V_{N^{\prime}}^{p}=\dfrac{V_{N}^{c}%
+1}{2}=\dfrac{2+V_{\varepsilon}+\tau V_{\text{\textrm{th}}}}{2}.
\end{equation}

The covariance between the input and output mode, for quadratures $q_{M}$ and
$q_{B^{\prime}}$, is given by%
\begin{equation}
\mathrm{Cov}(q_{M},q_{B^{\prime}})=\sqrt{\frac{\tau}{2}}V_{M}.
\end{equation}
We can build the following statistical estimator%
\begin{equation}
\hat{\sigma}_{MB^{\prime}}=\frac{1}{m}\sum_{i=1}^{m}M_{q,i}B_{q,i}^{\prime},
\end{equation}
compute its expectation value, obtaining%
\begin{equation}
\mathbb{E}\left[  \hat{\sigma}_{MB^{\prime}}\right]  =\sqrt{\frac{\tau}{2}%
}V_{M},
\end{equation}
and the variance%
\begin{align}
V_{\text{\textrm{cov}}}^{q}  &  =\frac{1}{m^{2}}\sum_{i=1}^{m}%
\text{\textrm{Var}}(M_{q,i}B_{q,i}^{\prime}),\nonumber\\
&  =\frac{\tau V_{M}^{2}+V_{M}V_{N^{\prime}}^{q}}{m},\nonumber\\
&  =\frac{\tau V_{M}^{2}}{2m}\left(  2+\frac{V_{N}^{q}+1}{\tau V_{M}}\right)
.
\end{align}
Then, assuming that $q_{M}$ and $q_{N^{\prime}}$ are independent variables,
with zero mean, we obtain the following expression for the estimator of the
transmissivity
\begin{equation}
\hat{\tau}=\frac{2\hat{\sigma}_{MB^{\prime}}^{2}}{V_{M}^{2}}=\frac
{2V_{\text{\textrm{cov}}}^{q}}{V_{M}^{2}}\left(  \frac{\hat{\sigma
}_{MB^{\prime}}}{\sqrt{V_{\text{\textrm{cov}}}^{q}}}\right)  ^{2},
\end{equation}
where $\left(  \frac{\hat{\sigma}_{MB^{\prime}}}{\sqrt{V_{\text{\textrm{cov}}%
}^{q}}}\right)  ^{2}$ is chi-squared distributed. Therefore, the expectation
value is given by
\begin{align}
\mathbb{E}(\hat{\tau})  &  =\frac{2V_{\text{\textrm{cov}}}^{q}}{V_{M}^{2}%
}\left(  1+\frac{\hat{\sigma}_{MB^{\prime}}^{2}}{V_{\text{\textrm{cov}}}^{q}%
}\right)  ,\nonumber\\
&  =\frac{2\hat{\sigma}_{MB^{\prime}}^{2}}{V_{M}^{2}}+\mathcal{O}%
(1/m),\nonumber\\
&  =\tau+\mathcal{O}(1/m)
\end{align}
and the variance
\begin{align}
\mathrm{Var}(\hat{\tau})  &  =\frac{8(V_{\text{\textrm{cov}}}^{q})^{2}}%
{V_{M}^{4}}\left(  1+2\frac{\hat{\sigma}_{MB^{\prime}}^{2}}%
{V_{\text{\textrm{cov}}}^{q}}\right)  ,\nonumber\\
&  =\frac{\frac{16\tau^{2}V_{M}^{4}}{4m}\left(  2+\frac{V_{N}^{q}+1}{\tau
V_{M}}\right)  }{V_{M}^{4}}+\mathcal{O}(1/m^{2}),\nonumber\\
&  =\frac{4\tau^{2}}{m}\left(  2+\frac{V_{N}^{q}+1}{\tau V_{M}}\right)
+\mathcal{O}(1/m^{2}).
\end{align}
For $m\gg1$, we neglect terms proportional to $1/m^{2}$ and write the variance
of $\hat{\tau}$ as follows%
\begin{equation}
\sigma_{q}^{2}:=\frac{4\tau^{2}}{m}\left(  2+\frac{V_{N}^{q}+1}{\tau V_{M}%
}\right)  .
\end{equation}
It is clear that repeating these steps for quadrature $\hat{p}_{B}$, we get
\begin{equation}
\sigma_{p}^{2}:=\frac{4\tau^{2}}{m}\left(  2+\frac{V_{N}^{p}+1}{\tau V_{M}%
}\right)  ,
\end{equation}
where the difference from $\sigma_{q}^{2}$ is the squeezing term of $V_{N}%
^{p}$.

From these, one can compute the optimal linear combination given by%
\begin{equation}
\sigma_{\text{\textrm{nsw}}}^{2}=\frac{1}{\sigma_{q}^{-2}+\sigma_{p}^{-2}}.
\end{equation}
Assuming that Alice starts the preparation from coherent states, we have that
$V_{s}=1$ and $\sigma_{\mathrm{c,}\text{\textrm{nsw}}}^{2}$ has the simpler
form%
\begin{equation}
\sigma_{\text{\textrm{c,nsw}}}^{2}=\frac{2\tau^{2}}{m}\left(  2+\frac
{V_{N}^{c}+1}{\tau V_{M}}\right)  .
\end{equation}

By solving Eq.~(\ref{eq:noisevarnswq}) with respect to $V_{\varepsilon}$ and
using the estimators of $V_{N^{\prime}}^{q}$ and $\tau$, we obtain%
\begin{equation}
\hat{V}_{\varepsilon}=2\hat{V}_{N^{\prime}}^{q}-\hat{\tau}%
(V_{\text{\textrm{th}}}+V_{s}-1)-2.
\end{equation}
We can replace the expression for $\hat{V}_{N^{\prime}}^{q}$ with%
\begin{equation}
\hat{V}_{N^{\prime}}^{q}=\frac{V_{N^{\prime}}^{q}}{m}\sum_{i=1}^{m}\left(
\frac{B_{q,i}-\sqrt{\frac{\tau}{2}}M_{q,i}}{\sqrt{V_{N^{\prime}}^{q}}}\right)
^{2}%
\end{equation}
which is chi-squared distributed, with mean $m$ and variance $2m$, because
$\left(  B_{i}^{q}-\sqrt{\frac{\tau}{2}}M_{i}\right)  /\sqrt{V_{N^{\prime}%
}^{q}}$ is a linear combination of normally distributed variables, having unit
variance and zero mean. Therefore, we obtain the following mean value for the
excess noise%
\begin{equation}
\mathbb{E}(\hat{V}_{\varepsilon})=\mathbb{E}\left(  2\hat{V}_{N^{\prime}}%
^{q}-\hat{\tau}(V_{\text{\textrm{th}}}+V_{s}-1)-2\right)  :=V_{\varepsilon}%
\end{equation}
and its variance, which is given by%
\begin{align}
s_{q}^{2}  &  :=\text{\textrm{Var}}(\hat{V}_{\varepsilon})=\text{\textrm{Var}%
}\left(  2\hat{V}_{N^{\prime}}^{q}-\hat{\tau}(V_{\text{\textrm{th}}}%
+V_{s}-1)-2\right) \nonumber\\
&  =\frac{4}{m^{2}}(V_{N^{\prime}}^{q})^{2}2m+(V_{\text{\textrm{th}}}%
+V_{s}-1)^{2}\sigma_{\text{\textrm{nsw}}}^{2}\nonumber\\
&  =\frac{2}{m}(V_{N}^{q}+1)^{2}+(V_{\text{\textrm{th}}}+V_{s}-1)^{2}%
\sigma_{\text{\textrm{nsw}}}^{2}. \label{sq}%
\end{align}
The same steps provide the expression of the variance for quadrature $p_{B}$
which is%
\begin{equation}
s_{p}^{2}=\frac{2}{m}(V_{N}^{p}+1)^{2}+(V_{\text{\textrm{th}}}+1/V_{s}%
-1)^{2}\sigma_{\text{\textrm{nsw}}}^{2}, \label{sp}%
\end{equation}
and from Eq.~(\ref{sq}) and Eq.~(\ref{sp}), we obtain%
\begin{equation}
s_{\text{\textrm{nsw}}}^{2}=\frac{1}{s_{q}^{-2}+s_{p}^{-2}},
\end{equation}
which for $V_{s}=1$ simplifies to%
\begin{equation}
s_{\text{\textrm{c,nsw}}}^{2}=\frac{\left(  V_{N}^{c}+1\right)  ^{2}}{m}%
+\frac{V_{\text{\textrm{th}}}^{2}\sigma_{\mathrm{c}\text{\textrm{,nsw}}}^{2}%
}{2}.
\end{equation}
Now, assuming the general case of moderately squeezed initial states, we can
write the confidence intervals which are taken~\cite{ruppertPRA} as follows%
\begin{align}
\tau^{\text{\textrm{low}}}=  &  \hat{\tau}-6.5\sigma_{\text{\textrm{nsw}}%
}(\hat{\tau},\hat{V}_{\varepsilon}),\\
V_{\varepsilon}^{\text{\textrm{up}}}=  &  \hat{V}_{\varepsilon}%
+6.5s_{\text{\textrm{nsw}}}(\hat{\tau},\hat{V}_{\varepsilon}),
\end{align}
assuming an error probability for the parameter estimation of the order of
$\varepsilon_{PE}=10^{-10}$.

Finally, proceeding as in
Sec.~\ref{Rate at various electromagnetic wavelengths}, we can write a key
rate of the form%
\begin{equation}
\label{nsw-rate}\tilde{K}=(1-r)\left[  \tilde{R}_{\xi}(\xi,V_{s}%
,V_{M},V_{\text{\textrm{th}}},V_{\varepsilon}^{\text{\textrm{up}}}%
,\tau^{\text{\textrm{low}}})-\Delta\right]  ,
\end{equation}
where the rate $\tilde{R}_{\xi}$ given by the following expression%
\begin{equation}
\tilde{R}_{\xi}=\xi\tilde{I}_{AB}-\tilde{\chi},
\end{equation}
where%
\begin{equation}
\tilde{I}_{AB}=\log_{2}\left[  1+\frac{\tau V_{M}}{2+V_{\varepsilon}+\tau
V_{\mathrm{th}}}\right]  ,
\end{equation}
and the expression of the Holevo function $\tilde{\chi}$ depends on the
implementation of the no-switching protocol, i.e., if the parties use direct
or reverse reconciliation.


\end{document}